\documentclass{aa}  

\usepackage{lipsum}
\usepackage{graphicx}
\usepackage{txfonts}
\usepackage{xcolor}
\usepackage{comment}
\usepackage{gensymb}

\begin{document}

   \title{\textsc{PIMMS}: Pulsation-Informed Magnetic Mapping of Stars with Zeeman-Doppler Imaging}
   \subtitle{I. Formalism and numerical tests}

   \author{C. Gutteridge\inst{1} \and 
           C. Neiner\inst{1} \and 
           C. Catala\inst{1} \and
           C. P. Folsom\inst{2}}

   \institute{LIRA, Observatoire de Paris, PSL University, CNRS, Sorbonne Universit\'e, Universit\'e Paris Cit\'e, CY Cergy University, 5 place Jules Janssen, 92195 Meudon, France\\
   \email{chloe.gutteridge@obspm.fr}
   \and
   Tartu Observatory, University of Tartu, Observatooriumi 1, T\~{o}ravere, 61602, Estonia }

   \date{Received <date>; accepted <date>}

% \abstract{}{}{}{}{} 
% 5 {} token are mandatory
 
  \abstract
   {
   Magneto-asteroseismology is a novel technique allowing for more precise determinations of internal properties of magnetic pulsating stars, but requires an accurate characterisation of the surface magnetic field, not previously possible with Zeeman-Doppler Imaging (ZDI) due to the time-dependent surface velocity of pulsating stars. 
   }
   {We aim to develop a new version of ZDI, which creates an additional surface velocity map, that includes the time-dependent velocities of surface elements due to pulsations.}
   {We present a new code, \textsc{PIMMS}: Pulsation-Informed Magnetic Mapping of Stars, which uses a surface-integrated line profile model that accounts for the additional Doppler shifts of local lines caused by pulsations. It is then possible to fit this model to spectropolarimetric observations, reconstructing maps of the surface brightness, magnetic field, and the velocity field due to the combination of pulsation and rotation. In this paper, we present and test \textsc{PIMMS} extensively, to understand its limitations and data requirements.}
   {We find that \textsc{PIMMS} can accurately reproduce the magnetic fields and brightness distributions of realistic models of pulsating hot stars. The required number of observations is higher than that required for ZDI of a non-pulsating star due to the additional velocity map that must be disentangled from surface brightness variations. \textsc{PIMMS} is now ready to be applied to real stars.}
   {}


   \maketitle
%
%-------------------------------------------------------------------

\section{Introduction}

\begin{figure*}[t!]
\centering
   \includegraphics[width=17cm]{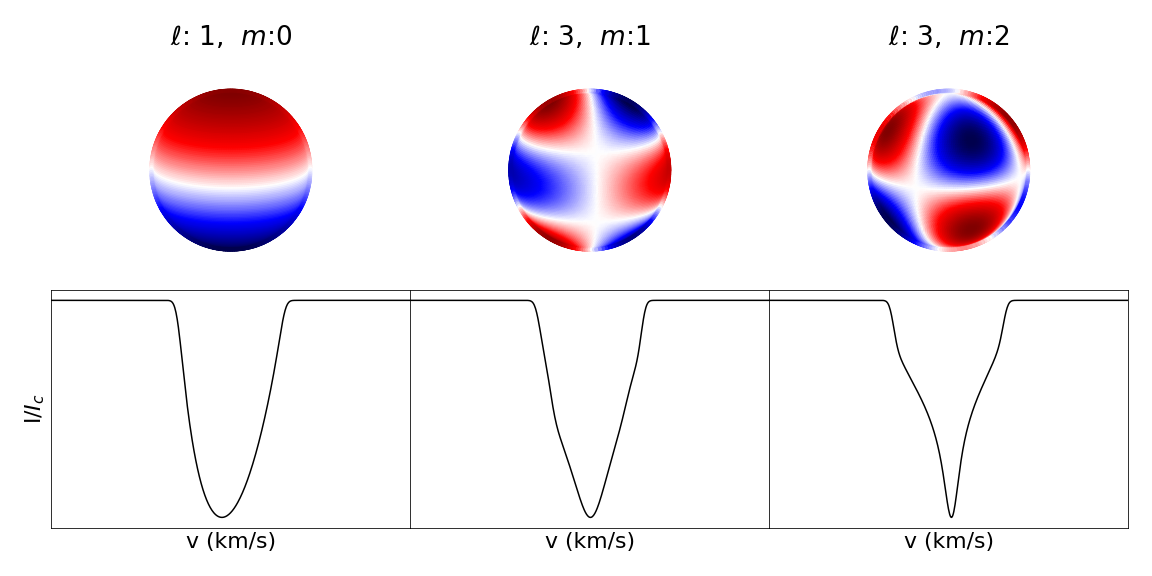}
     \caption{Examples of line profiles for stars with three different pulsation modes, all with inclination angles of 75$\degree$.}
     \label{line_profile_example}
\end{figure*}

Zeeman-Doppler Imaging (ZDI; \citealt{ZDI_original}) is a tomographic technique used to map the magnetic fields of stars using spectropolarimetric data. The Doppler shifts of stellar lines are dependent on the surface velocity, which is in turn dependent on the latitude of the source relative to the rotation axis, so can therefore be used to probe positions. The Zeeman effect reveals the magnitude of the surface magnetic field strength at these positions through the size of the frequency separation of the split stellar lines, and orientation of the magnetic field through polarisation. ZDI is often performed with circularly polarised data (Stokes V) as the Zeeman signature in V is $\sim$10 times stronger than in linearly polarised data (Stokes Q and U) for most stars. At the cost of extra observation time, ZDI can be performed with all Stokes parameters, which provides stronger constraints on the magnetic field, and remove degeneracies in some cases \citep{ZDI_QU}.

Many codes exist to reconstruct stellar magnetic fields, from the original work of \cite{ZDI_Brown}, built upon by \cite{crosstalk}, to \textsc{INVERS13} \citep{INVERS13}, and more recently \textsc{ZDIpy} \citep{ZDIpy}. All of these codes work on three assumptions: the star can be modelled as a sphere, the velocity at any point on the stellar surface is only dependent on position, and the only variation in observed magnetic field strength and surface brightness is due to rotation. Hence, subsequent work has focused on developing codes to account for stars for which these assumptions do not apply: the modification of \textsc{ZDIpy} by \cite{ZDIpy_oblate} allows for the oblate geometries of rapid rotators, while the \textsc{TIMeS} code \citep{TIMeS} can map magnetic fields that vary on timescales less than the observing period. An attempt has previously been made to allow for Zeeman-Doppler imaging of stars with intrinsic variability in their surface brightness by \cite{ZDI_var}, using a simple model of linear variations in logarithmic relative surface brightness with time. This however was found only to provide minor improvements on standard ZDI as it could not fully represent the observed variability in surface features.

Zeeman-Doppler Imaging has not been reliably performed for pulsating stars previously, as the stellar surface velocity is not solely dependent on rotation: a key assumption of all previous codes. Pulsating stars show variations in their line profiles due to the Doppler shifts caused by pulsation velocities (Fig. \ref{line_profile_example}). In this paper we present the first ZDI code developed in order to account for the time-dependent surface velocity of pulsating stars, \textsc{PIMMS}: Pulsation-Informed Magnetic Mapping of Stars. Being able to accurately reconstruct magnetic field strengths and geometries from spectropolarimetric data will lead to more progress in the emerging field of magneto-asteroseismology, where asteroseismology is performed with the additional consideration of the surface and interior magnetic fields (\citealt{magentoasteroseismology_Fuller}; \citealt{magentoasteroseismology_Li}; \citealt{magnetoasteroseismology}). Magnetic fields modify the pulsation modes and play an important role in the dynamics of the stellar interior, e.g. by decreasing or suppressing internal differential rotation. These effects can be inferred if a surface magnetic field is detected and then taken into account in the modelling.

Due to a lack of targets with both sufficient spectropolarimetric data, and rich and well-defined pulsations, magneto-asteroseismology has so far been attempted for only three hot stars: two $\beta$~Cep stars, $\beta$~Cep \citep{betaCep} and V2052~Oph \citep{V2052Oph}; and the SPB star HD~43317 (\citealt{HD43317}; \citealt{HD43317_second_paper}). In the case of $\beta$~Cep, the wrong rotation period was used in the model resulting in inaccurate results \citep{beta_cep_rot}, while for V2052~Oph, the magnetic field was only used to explain the seismic results rather than included in the model itself. Only the model for HD~43317 is relatively complete, but for all three stars the magnetic field strength and geometry was determined without the precision that ZDI can provide. 

In Sect. 2 we describe the stellar pulsation model and subsequent line profile synthesis we have implemented into \textsc{ZDIpy} to create \textsc{PIMMS}. We then extensively test \textsc{PIMMS} in Sect. 3 with models of realistic hot magnetic pulsators. In Sect 4. we examine factors that must be considered when applying \textsc{PIMMS} to real stars, then we discuss our results and conclusions in Sect. 5.
%--------------------------------------------------------------------
\section{Stellar pulsation model}

\subsection{Initialisation file}\label{init_file}

Like \textsc{ZDIpy}, \textsc{PIMMS} requires certain variables to be assigned values prior to running, in an initialisation file. These include: the stellar parameters inclination $i$, limb darkening coefficient $\eta$, projected equatorial velocity $v_\text{eq}\sin i$, and rotation period, $P = 2\pi/\Omega$; the pulsation parameters degree $\ell$, azimuthal order $m$, and frequency $f$; and the model parameters resolution (number of surface elements), velocity range of the line profiles to be considered when fitting, and the Julian date for the start of rotation cycle 0. The file names are also provided here, alongside their Julian dates of observation, and central velocity corrections for line profiles due to barycentric and binary motion if present. This is important as \textsc{PIMMS} only considers velocity variations due to pulsation, and would therefore try to fit these external velocity variations as such. 

This file is also where fitting parameters can be controlled. \textsc{PIMMS} uses the maximum entropy image reconstruction algorithm from \textsc{ZDIpy}. Target values can be provided for both the $C_\text{aim}$ and `TEST' criteria \citep{max_entropy}. $C_\text{aim}$ controls the balance between maximum entropy and $\chi^2$ in the final solution, while `TEST' provides the convergence criterion of the model. Each map (pulsation/velocity, brightness, and magnetic) can be fitted individually or in any combination with one another, and those maps can then be loaded as constants or starting points for subsequent fits, which significantly improves the goodness of fit. For \textsc{ZDIpy}, which assumes a surface velocity constant in time, the suggested running order is to initially do a fit of the brightness map, then use that brightness map to fit the magnetic field map, then use both maps as a starting point for a final fit, where both are allowed to vary. For \textsc{PIMMS}, which additionally fits a time-dependent velocity map with pulsations, we suggest the following running order:
\begin{enumerate}
    \item Fit only the pulsations (assuming a surface brightness uniform in space and time)
    \item Using the fitted pulsation parameters, fit the surface brightness
    \item Fit both pulsation and brightness, using the results from the previous two fits as initial parameters
    \item Keep the pulsation parameters and surface brightness map from the previous fit constant while fitting the magnetic field
    \item Use the results of the previous fit as the starting point for a final fit that allows all pulsation, brightness and magnetic field parameters to vary
\end{enumerate}

The allowed field geometries can also be controlled. \textsc{PIMMS} uses the exact same spherical harmonic description of the magnetic field and fitting methods as in \textsc{ZDIpy}, which follow \cite{magnetic_field_desc}. The constants $\alpha_{\ell m}$, $\beta_{\ell m}$, and $\gamma_{\ell m}$ describe the contributions of the radial field, the azimuthal and meridional terms of the poloidal field, and the azimuthal and meridional terms of the toroidal field respectively, for each spherical harmonic component of degree $\ell$ and order $m$. Fitting only certain constants or constraining the allowed values will result in different field structures: fitting $\alpha$ and $\beta$, and setting $\gamma = 0$ provides a poloidal field; fitting $\alpha=\beta$, and setting $\gamma = 0$ provides a potential field; fitting $\alpha=\beta$ and $\gamma$ provides a potential-toroidal field. Fitting all three unconstrained allows for an arbitrary vector field. The complexity of the allowed magnetic field is controlled by the number of spherical harmonic components fitted. For a maximal spherical harmonic degree $\ell_\text{max}$, there are $N_\text{sph} = \frac{1}{2}\ell_{max}(\ell_{max}+3)$ components. $\ell_\text{max}$ is provided in the initialisation file and should be carefully set when performing ZDI with either \textsc{ZDIpy} or \textsc{PIMMS} to ensure sufficient complexity to accurately describe the field.

\subsection{Stellar surface grid}

We begin by defining a unit sphere, on a system of Cartesian axes, that the stellar model will be described by:
$$
      x = \sin{\theta}\cos{\phi}   \, , \;
      y = \sin{\theta}\sin{\phi}   \, , \;
      z = \cos{\theta}
$$
where polar angle $\theta$ is the angle from the $Z$ axis to the radial line from the origin to a given point $(x,y,z)$, while the azimuthal angle $\phi$ is the angle between the $X$ axis and projection of the point onto the $X-Y$ plane.

We then separate the unit sphere into surface area elements of approximately equal area $A$, calculated by Eq.~\ref{element_area}. We do this first by separating the sphere into rings of equal latitude ranges, and calculating the surface area of each ring, $A_{\theta_1 \rightarrow \theta_2}$, using Eq.~\ref{ring_area}. We then allocate a number of elements proportional to the fraction of the total surface area each ring contains, equally space them in longitude, and calculate the true areas of each of these surface elements.

\begin{align}
    \label{element_area}
    A &= \iint \sin{\theta}\,d\theta\,d\phi \\
    \label{ring_area}
    A_{\theta_1 \rightarrow \theta_2} &= 2\pi|\cos\theta_1 - \cos\theta_2| 
\end{align}

We first assert that the rotation axis, which we choose to be the $Z$ axis, is perpendicular to the line of sight to the star. We will need to transform this model to account for the inclination angle of the star's rotation and pulsation axis to the line of sight. In \textsc{PIMMS} we assume these to be the same, as is the case in most stars. Notable exceptions include roAp stars \citep{roAp} and stars with tidally-induced pulsations \citep{tidal}, the latter of which is non-spherical and therefore would not be well represented by our model regardless. The rotation matrix about the Cartesian axes can be defined as:

\begin{equation}
    \label{matrix_transform}
    R = 
    \begin{pmatrix}
        \cos(b)\cos(c) & 
        \sin(a)\sin(b)\cos(c)  &
        \cos(a)\sin(b)\cos(c) \\
        &- \cos(a)\sin(c) & + \sin(a)\sin(c)\\
        &&\\
        \cos(b)\sin(c) & 
        \sin(a)\sin(b)\sin(c) & 
        \cos(a)\sin(b)\sin(c) \\
        &+ \cos(a)\cos(c)&- \sin(a)\cos(c)\\
        &&\\
        -\sin(b) & \sin(a)\cos(b) & \cos(a)\cos(b) 
    \end{pmatrix}
\end{equation}
where $a$, $b$ and $c$ are the angles about the $X$, $Y$ and $Z$ axes respectively.

To rotate the model star, we apply a transformation using $ c = \Omega t$, where $\Omega$ is the angular rotation frequency of the star and $t$ is the time past since reference time 0, e.g. Eq.~\ref{rot_transform}.
\begin{equation} 
    \label{rot_transform}
    R_\Omega = 
    \begin{pmatrix}
        \cos(\Omega t) &  -\sin(\Omega t) & 0 \\ 
        \sin(\Omega t) &   \cos(\Omega t) & 0 \\ 
              0 &         0 & 1 
    \end{pmatrix}
\end{equation}

To incline the star, we set $a$ to the inclination angle $i$ provided in the initialisation file.
\begin{equation}
    \label{incl_transform}
    R_i = \begin{pmatrix}1 & 0 & 0 \\ 0  & \cos(i) & - \sin(i) \\ 0 & \sin(i) & \cos(i) \end{pmatrix}
\end{equation}

\subsection{Stokes profiles}\label{stokes_sec}

We follow the derivation of Eq. 6.49 in \cite{Aerts2010} to describe the observed line profile. The line forming region is assumed to be geometrically thin, so that the dependence of velocity and magnetism on line depth can be ignored. Each local line profile $p_{ij}$, for surface element $(i,j)$, can be approximated as a normalised Gaussian. $i$ and $j$ are the indexes in colatitude $\theta$ and longitude $\phi$ respectively. The variance, $\sigma$, controls the intrinsic line width, which is affected by natural, pressure, and thermal broadening. 
\begin{equation}
    p_{ij}(v,t) = \frac{1}{\sqrt{2\pi}\sigma}\exp\left(-\frac{(v_{ij}(t)-v)^2}{2\sigma^2}\right)
\end{equation}
Each local line will be Doppler shifted by an amount dependent on the velocity of its corresponding area element, giving the new line centre on the velocity axis, $v_{ij}$. This velocity is dependent on both rotation and pulsation velocities, the latter of which is dependent on time, $t$. Eq. \ref{velocity_eq} defines the total velocity along the line of sight to the observer. $\theta'_i$ and $\phi'_j$ are the coordinates $\theta_i$ and $\phi_j$ transformed such that the polar axis is inclined by stellar inclination angle $i$, while the zero point of the azimuthal angles remains the same. 

The pulsation velocities in the spherical coordinate system are defined as follows,
\begin{equation*}
    \begin{pmatrix}
        v_{r}\\
        v_{\theta}\\
        v_{\phi}
    \end{pmatrix}_\text{puls}
    =
    \begin{pmatrix}
        \Xi_{nl}(r)P^m_l(\cos\theta)\sin(m\phi-\omega t-\varphi)\\
        \\
        \Theta_{nl}(r)\frac{dP^m_l(\cos\theta)}{d\theta}\sin(m\phi-\omega t-\varphi)\\
        \\
        \Theta_{nl}(r)\frac{m}{\sin\theta}P^m_l(\cos\theta)\cos(m\phi-\omega t-\varphi)
    \end{pmatrix}_\text{puls}
\end{equation*}

and are transformed to Cartesian coordinates by matrix $\mathbf{S}(\theta_i,\phi_j)$. The velocity along the line of sight is defined as $v_{z,ij}$.

\begin{equation}
    \label{velocity_eq}
    \begin{pmatrix}
        v_{x,ij}\\
        v_{y,ij}\\
        v_{z,ij}
    \end{pmatrix}_\text{puls}
    =
    \mathbf{R}_i\mathbf{R}_\Omega \mathbf{S}(\theta_i,\phi_j)
    \begin{pmatrix}
        v_{r,ij}\\
        v_{\theta,ij}\\
        v_{\phi,ij}
    \end{pmatrix}_\text{puls}
\end{equation}
\begin{equation*}
    \mathbf{S}(\theta_i,\phi_j) = 
    \begin{pmatrix}
        \sin\theta_i\cos\phi_j&\cos\theta_i\cos\phi_j&-\sin\phi_j\\
        \sin\theta_i\cos\phi_j&\cos\theta_i\cos\phi_j&\cos\phi_j\\
        \sin\theta_i\cos\phi_j&\cos\theta_i\cos\phi_j&0\\
    \end{pmatrix}
\end{equation*}

The pulsation parameters degree $\ell \geq 0$, azimuthal order $m \in [-\ell,+\ell]$, and frequency $f$ are provided for each pulsation mode, along with inclination angle $i$, and equatorial velocity $v_\Omega$ for the star. The following pulsation variables are fitted: tangential to radial displacement ratio $K$, pulsation velocity $v_p$, and phase $\varphi$ at time $t=0$ defined in the initialisation file. $P^m_\ell(\cos\theta)$ are Legendre functions given for $m \geq 0$ by
\begin{equation}
    \label{legendre_pos}
    P^m_\ell(\cos\theta)=\frac{1}{2^\ell\ell!}\left(1-\cos^2\theta\right)^{m/2}\frac{d^{\ell+m}}{d\cos^{\ell+m}\theta}\left(\cos^2\theta - 1\right)^\ell
\end{equation}
We remove the differential factor in the Legendre function, to provide an analytical solution for use in \textsc{PIMMS}, using the binomial theorem expansion, and the substitution $\vartheta = \cos{\theta}$. 
\begin{align*}
    P^m_\ell(\vartheta) &= \frac{1}{2^\ell\ell!}\left(1-\vartheta^2\right)^{m/2}\frac{d^{\ell+m}}{d\vartheta^{\ell+m}}\left(\vartheta^2 - 1\right)^\ell \\
    &= \frac{1}{2^\ell\ell!}\left(1-\vartheta^2\right)^{m/2}\frac{d^{\ell+m}}{d\vartheta^{\ell+m}}\left[\sum^\ell_{k=0}(-1)^k\frac{\ell!}{k!(\ell-k)!}\vartheta^{2(\ell-k)}\right]\\
    &= \frac{1}{2^\ell}\left(1-\vartheta^2\right)^{m/2}\sum^{k\leq\frac{\ell-m}{2}}_{k=0}\frac{(-1)^k}{k!(\ell-k)!}\frac{[2(\ell-k)]!}{[\ell-2k-m)]!}\vartheta^{\ell-2k-m}
\end{align*}
Reversing the $\vartheta = \cos{\theta}$ substitution, and introducing a function $\kappa(\ell, m, k)$ dependent only on the mode parameters, we get the form used in the code to compute the Legendre functions. 
\begin{equation}
    \label{legendre_exp}
    P^m_\ell(\cos\theta) = \sin^m\theta \sum^{k\leq\frac{\ell-m}{2}}_{k=0}\kappa(\ell, m, k)\cos^{\ell-2k-m}\theta
\end{equation}
where $\kappa(\ell, m, k)$ is defined as:
\begin{equation}
    \label{alpha}
    \kappa(\ell, m, k) = \frac{1}{2^\ell}\frac{(-1)^k}{k!(\ell-k)!}\frac{[2(\ell-k)]!}{[\ell-2k-m]!}
\end{equation}
We then need to differentiate Eq. \ref{legendre_exp} by $\theta$ for the differential term in the sum of Eq. \ref{velocity_eq}.
\begin{align*}
    \frac{d P^m_\ell(\cos\theta)}{d\theta} &= \frac{d\sin^m\theta}{d\theta}\sum^{k\leq\frac{\ell-m}{2}}_{k=0}\kappa(\ell, m, k)\cos^{\ell-2k-m}\theta\\
    &\quad   +\sin^m\theta\sum^{k\leq\frac{\ell-m}{2}}_{k=0}\kappa(\ell, m, k)\frac{d \cos^{\ell-2k-m}\theta}{d\theta} \\
    &= \sin^{m-1}\theta\sum^{k\leq\frac{\ell-m}{2}}_{k=0}\kappa(\ell, m, k)\cos^{\ell-2k-m-1}\theta\\
    &\quad \times\left[m + (2k-\ell)\sin^2{\theta}\right]
\end{align*}
For $m < 0$, a proportionality relationship can be used:
\begin{equation}
    \label{legendre_neg}
    P^{-m}_\ell(\cos\theta)=(-1)^m\frac{(\ell-m)!}{(\ell+m)!}P^m_\ell(\cos\theta)
\end{equation}
As Eq.~\ref{legendre_neg} is linear, the constant factors can also be used to give the differentiated associate Legendre polynomials with $m < 0$, i.e.
\begin{equation*}
    \frac{d P^{-m}_\ell(\cos\theta)}{d\theta} =(-1)^m\frac{(\ell-m)!}{(\ell+m)!}\frac{d P^m_\ell(\cos\theta)}{d\theta} 
\end{equation*}

Each local line profile is weighted in the sum by the brightness of the elemental area associated. The weights are calculated from product of: the continuum intensity at $(\theta_i, \phi_j)$, $I_0$; the projected area of the element, $\sin\theta'_i\cos\theta'_i\,\Delta\theta_i'\,\Delta\phi'_j$; and the limb darkening coefficient $h_\lambda(\theta_i')$, from the linear limb darkening law $h_\lambda(\theta_i') = 1 - \eta + \eta\cos(\theta_i')$ \citep{limb_darkening}. This is the same limb darkening law used in the \textsc{ZDIpy}. The sum is taken over the visible stellar surface, i.e. $\theta' \in [0,\,\frac{\pi}{2}]$, $\phi' \in [0,\,2\pi]$, and renormalised by the sum of the weights. Eq. \ref{stokes_I_eq} is equivalent to the Stokes I profile.
\begin{equation}
    \label{stokes_I_eq}
    p(v,t) = 1- \frac{\sum_{i,j}I_0(\theta_i, \phi_j)h_\lambda(\theta'_i)p_{ij}(v,t)\sin\theta'_i\cos\theta'_i\,\Delta\theta_i'\,\Delta\phi'_j}{\sum_{i,j}I_0(\theta_i, \phi_j)h_\lambda(\theta'_i)\sin\theta'_i\cos\theta'_i\,\Delta\theta_i'\,\Delta\phi'_j}
\end{equation}

The magnetic model remains unchanged from \textsc{ZDIpy}, albeit with the Stokes V profiles now calculated from the derivative of the new line profile model. 
\begin{equation}
    \label{stokes_v_eq}
    V_{ij}(\lambda) = g_\text{eff}\frac{\lambda_0^2e}{4\pi m_e c}B_{l,ij}\frac{dp_{ij}(v,t)}{dv}\frac{dv}{d\lambda}
\end{equation}

\begin{figure*}
\centering
\includegraphics[width=5.5cm]{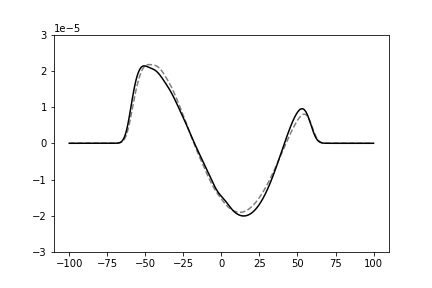}
\includegraphics[width=5.5cm]{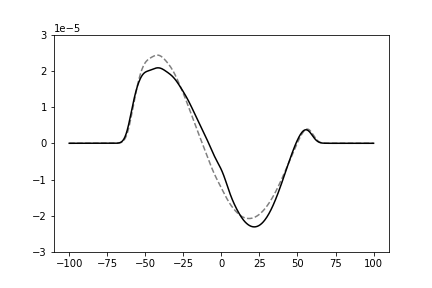}
\includegraphics[width=5.5cm]{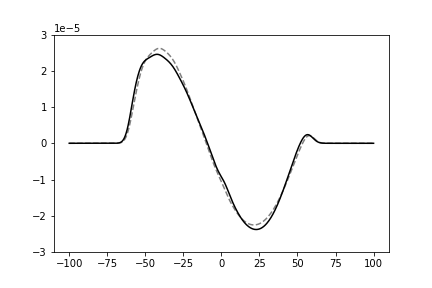}
     \caption{Model Stokes V signatures for stars at rotation phases 0, 0.5, and 1. The solid black line is for a star with a single $\ell=2$, $m=-2$ pulsation mode, while the dashed grey line is for a non-pulsating star. The stars have a uniform surface brightness, and identical rotation periods and magnetic fields. The model Stokes V signatures deviate strongly from each other, even at equivalent rotation phases i.e. 0 and 1, due to the pulsation velocities.}
     \label{stokes_V}
\end{figure*}

Examples of our model Stokes V profiles can be seen in Fig. \ref{stokes_V}. As the pulsations affect the Stokes V signature, the magnetic field model can inform our pulsation fit, which is why we recommend allowing all parameters to vary in the final step of running \textsc{PIMMS}.

\section{Tests on Simulations}\label{test}

For the following tests, we use \textsc{PIMMS} to generate the simulated line profiles, using approximately 10000 equally sized cells to represent the stellar surface. The model is inclined in rotation by $i=55\degree$ and the magnetic field is inclined to the rotation axis by $\beta=40\degree$. The star rotates with a frequency of $f_\text{rot}=1.10$~d$^{-1}$, corresponding to $v\sin i = 60$~km~s$^{-1}$, and has a local line width of $\sigma = 3$~km~s$^{-1}$. We follow the ideal fitting order described in \ref{init_file}, and allow fitting of magnetic field topologies up to spherical harmonic degree $\ell_{\text{max}} = 5$, though our test field is a simple dipole ($\ell_{\text{max}} = 1$). Our test pulsation mode parameters, representative of real pulsating stars, can be found in Table \ref{table:1 - test mode params}. These are the measured pulsation mode parameters of $\beta$ Ceph star V2052 Oph, from the paper of \cite{V2052Oph}. We add Gaussian noise to every model line profile to give signal-to-noise ratios (S/N) of 1900, which replicate ideal but realistic observations.

\begin{table}[h]
\caption{Pulsation mode parameters used to test \textsc{PIMMS}. They are the same as those present in the magnetic $\beta$ Ceph star V2052 Oph \citep{V2052Oph}.}              % title of Table
\label{table:1 - test mode params}      % is used to refer this table in the text
\centering                                      % used for centering table
\begin{tabular}{c c c c}          % centered columns (4 columns)
\hline\hline                        % inserts double horizontal lines
ID & $f$ & $\ell$ & $m$ \\    % table heading
 & $(d^{-1})$ &  &  \\    % table heading
\hline                                   % inserts single horizontal line
    $f_1$ & 7.14846 & 0 & 0 \\      % inserting body of the table
    $f_2$ & 7.75603 & 4 & 3 \\
    $f_3$ & 6.82308 & 4 & 2 \\
\hline                                             %inserts single line
\end{tabular}
\end{table}

\subsection{Star with a dipolar magnetic field and a uniform surface brightness}\label{sect: uniform_brightness_test}

\begin{figure*}
\centering
   \includegraphics[width=5.5cm]{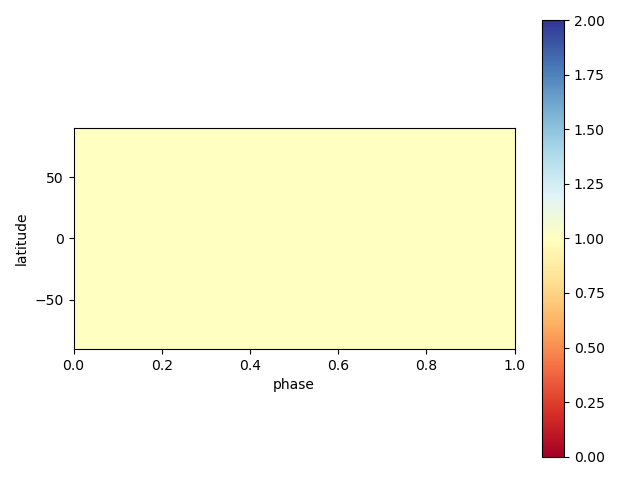}
   \includegraphics[width=5.5cm]{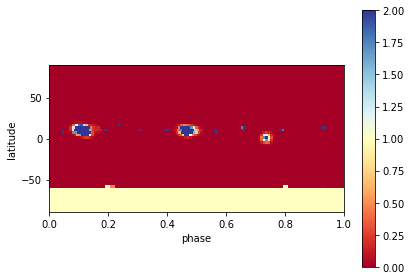}
   \includegraphics[width=5.5cm]{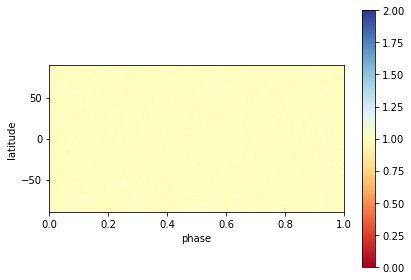}
     \caption{Test for a dipolar star with uniform brightness. The left-hand plot is the brightness map used to generate the model line profiles we test the codes on, the middle is the best fit brightness map produced by \textsc{ZDIpy}, and the final plot is the best fit brightness map produced by \textsc{PIMMS}.}
     \label{uniform_brightness}
\end{figure*}

\begin{figure*}
\centering
   \includegraphics[width=5.5cm]{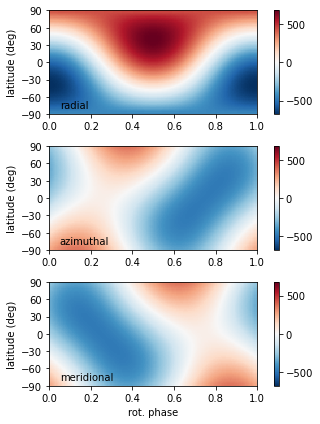}
   \includegraphics[width=5.5cm]{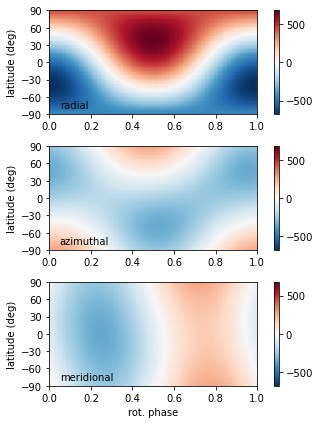}
   \includegraphics[width=5.5cm]{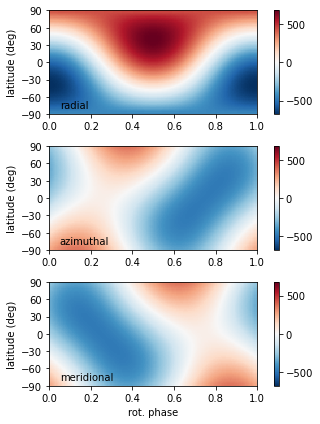}
     \caption{Test for a dipolar star with uniform brightness. The left-hand plot is the magnetic map used to generate the model line profiles we test the codes on, the middle is the outputted magnetic field map from \textsc{ZDIpy} assuming the input surface brightness map, and the final plot is the magnetic field map from \textsc{PIMMS} using the fitted brightness map.}
     \label{uniform_brightness_mag}
\end{figure*}

We begin by testing the ability to fit the pulsations of a star with a uniform surface brightness. We provide 100 observations equally spread over the rotation phase to ensure that the rotation phase coverage is not the limitation. The pulsation phase is well covered due to the comparatively short pulsation frequencies. \textsc{ZDIpy} fails to produce a physical brightness map, instead outputting a completely dark star with bright spots along a line of latitude around $+15\degree$ (see middle panel of Fig. \ref{uniform_brightness}). This angle corresponds to $i - \beta$. \textsc{ZDIpy} produces this wrong brightness map because it tries to compensate for the pulsations with spots, but as the rotation and pulsation frequencies are not at all similar, the resulting fitted line profiles do not resemble the simulated observations. We tested the ability of \textsc{ZDIpy} to fit the magnetic field assuming a uniform surface brightness, but even when the true surface brightness is provided, \textsc{ZDIpy} fails to reproduce the magnetic field, due to the incorrect attribution of velocities to position, resulting in additional weak poles being added (see middle panel of Fig. \ref{uniform_brightness_mag}). This confirms that \textsc{ZDIpy} should not be used for pulsating stars. 

\textsc{PIMMS} reproduces the uniform surface brightness correctly, with only very minor and evenly distributed variations due to the added noise (see right-hand panel of Fig. \ref{uniform_brightness}). The recovered magnetic field is essentially the same as the inputted field, with differences $<1\%$ (see right-hand panel of Fig. \ref{uniform_brightness_mag}). Using the inputted uniform surface brightness instead of the fitted surface brightness causes no significant difference in the model magnetic field when tested with a perfect dipole. 

\begin{table}
\caption{Summary of input and best fit values for the \textsc{ZDIpy} and \textsc{PIMMS} codes, when applied to 100 simulated observations of a $\beta$ Cep star with a dipolar magnetic field, and uniform surface brightness as described in Sect. \ref{sect: uniform_brightness_test}.}
\label{table:2 - uniform results} 
\centering
\begin{tabular}{c c c c}
\hline\hline                        
 Parameter &Input&\textsc{ZDIpy}&\textsc{PIMMS}\\
\hline                                   
    $v_{p,1}$ & 20.0000 & & 20.0217\\
    $\varphi_1$ & 0.0000 & & 0.0005\\
    $K_1$ & 0.0000 & & 0.0000\\
    $v_{p,2}$ & 0.5000 & & 0.4958\\
    $\varphi_2$ & 0.3000 & & 0.3001\\
    $K_2$ & 0.0100 & & 0.0109\\
    $v_{p,3}$ & 0.5000 & & 0.5088\\
    $\varphi_3$ & 0.7000 & & 0.7093\\
    $K_3$ & 0.1000 & & 0.0754\\
    $B_\text{d}$ & 583 & 523 & 581\\
\hline                                             %inserts single line
\end{tabular}
\end{table}

\subsection{Star with a dipolar magnetic field and a spotted surface}\label{spot_sec}

\begin{figure*}
\centering
   \includegraphics[width=5.5cm]{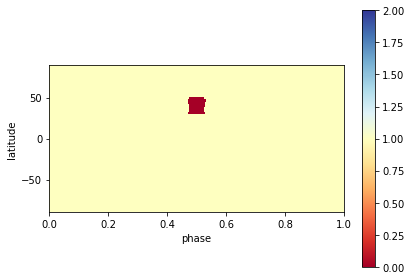}
   \includegraphics[width=5.5cm]{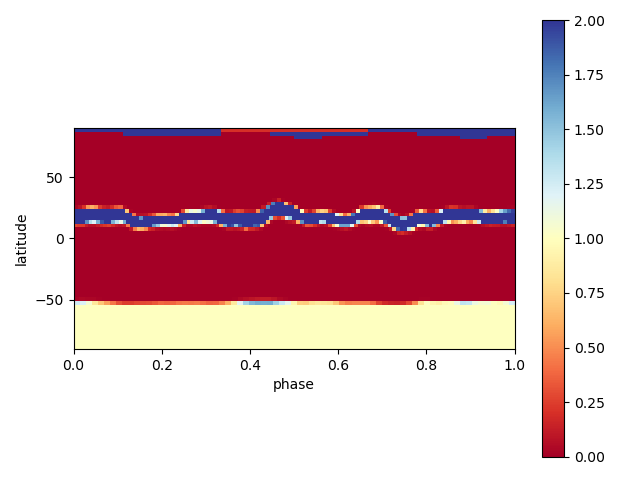}
   \includegraphics[width=5.5cm]{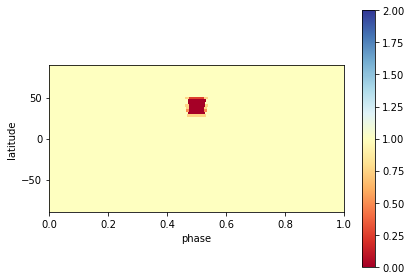}
     \caption{Same as Fig. \ref{uniform_brightness}, but for a star with a singular polar spot.}
     \label{spot_brightness}
\end{figure*}

\begin{figure*}
\centering
   \includegraphics[width=5.5cm]{img/dipole_in.png}
   \includegraphics[width=5.5cm]{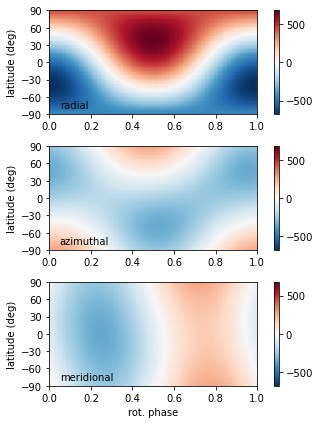}
   \includegraphics[width=5.5cm]{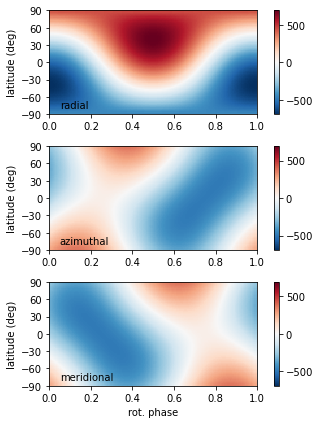}
     \caption{Same as Fig. \ref{uniform_brightness_mag}, but for a star with a singular polar spot.}
     \label{spot_brightness_mag}
\end{figure*}

We progress to a pulsating star with a single spot, aligned with the visible pole of the dipolar magnetic field. We again provide 100 observations equally spread over the rotation phase to ensure that the phase coverage is not the limitation. When applying \textsc{ZDIpy}, it results in an entirely dark star, except at the pole and again the line of latitude around $+15\degree$ which it considers bright (see middle panel of Fig. \ref{spot_brightness}). Once again, even when the true surface brightness is provided, \textsc{ZDIpy} fails to reproduce the magnetic field, with field strength being incorrectly attributed to higher order polar terms (see middle panel of Fig. \ref{spot_brightness_mag}).

In contrast, \textsc{PIMMS} produces the inputted polar spot, albeit smoothed out (see right-hand panel of Fig. \ref{spot_brightness}), because of the noise we added to the input data and the sampling of rotation phase. This smoothing causes slight overestimation of the field strength near the pole, but differences with regard to the inputted field are $<2\%$ (see right-hand panel of Fig. \ref{spot_brightness_mag}), and the position of the pole is not affected. This must nonetheless be taken seriously, as for the hot magnetic stars there are often temperature spots on the poles. The overestimation can be corrected by limiting the spherical harmonic degree $\ell_{\text{max}}$ to a lower value, but in a real case where we do not know the true topology, we could be losing resolution in structure and incorrectly attributing to increased strength in spots.

\begin{table}
\caption{Summary of input and best fit values for the \textsc{ZDIpy} and \textsc{PIMMS} codes, when applied to 100 simulated observations of a $\beta$ Cep star with a dipolar magnetic field, and a single spot on the magnetic pole as described in Sect. \ref{spot_sec}.}
\label{table:3 - spot results} 
\centering
\begin{tabular}{c c c c}
\hline\hline                        
 Parameter &Input&\textsc{ZDIpy}&\textsc{PIMMS}\\
\hline                                   
    $v_{p,1}$ & 20.0000 & & 20.0193\\
    $\varphi_1$ & 0.0000 & & 0.0005\\
    $K_1$ & 0.0000 & & 0.0004\\
    $v_{p,2}$ & 0.5000 & & 0.5074\\
    $\varphi_2$ & 0.3000 & & 0.3002\\
    $K_2$ & 0.0100 & & 0.0110\\
    $v_{p,3}$ & 0.5000 & & 0.5207\\
    $\varphi_3$ & 0.7000 & & 0.7088\\
    $K_3$ & 0.1000 & & 0.0971\\
    $B_\text{d}$ & 583 & 514 & 594\\
\hline                                             %inserts single line
\end{tabular}
\end{table}

\subsection{Required number of observations}\label{obs_req}

The number of observations required to reliably constrain the magnetic and pulsation model is dependent on many variables, e.g. number of pulsation modes, the mode geometries, number of chemical spots, magnetic field geometry, phase coverage distribution, and precision aim (see Sect. \ref{discussion} for more detail). It is therefore not possible to give a simple value here. In cases where a preliminary characterisation of the stellar magnetic field exists, and the pulsation modes have been previously identified, we advise modelling the star with these expected pulsation modes and magnetic field geometry in advance of observing, in order to estimate how many additional data must be obtained. Here we will demonstrate the data requirements for a model $\beta$~Cep star based on V2052~Oph, and a model SPB star with a single pulsation frequency close to its rotation frequency. Our $\beta$ Cep model has a dipolar field, spots on the magnetic poles (rather than the off-centre Helium poles in the case of V2052 Oph), and the three pulsation modes given in Table \ref{table:1 - test mode params}. This $\beta$ Cep model can be interpreted as the minimum data requirement for \textsc{PIMMS} to operate due to its relative simplicity. Our SPB model also has a dipolar field, with off-pole spots, and a single non-radial pulsation mode, ($\ell=2$, $m=0$, $f=1.15$~d$^{-1}$). We set the rotation frequency to $f_\text{rot}=1.10$~d$^{-1}$ to demonstrate a much more complicated case where the pulsation must be disentangled from permanent features, i.e. spots. In both cases, we aim for a fit threshold of $\chi^2<1.5$. 

First, we consider the case where we are able to schedule our observations such that we can perfectly distribute them over a single rotation cycle. In this case, we can recover the field strength and structure with a minimum of 19 observations for our model $\beta$ Cep pulsator. It should however be noted that mode identification would be difficult with so few observations. In cases where the pulsation modes have been well studied in pure spectroscopy, such that mode identification from spectropolarimetric observations is not required, equally spacing observations over the rotation phase could be an adequate observation strategy. However, the pulsation phase coverage must nonetheless still be considered, and will strongly affect the goodness of fit. The required number of observations is significantly higher for our model SPB at 109. Although there is only a single pulsation mode present, it needs to be disentangled from the line profile variations caused by spots, which is difficult over a single rotation period when the frequencies are so close. Although it converges with an ideal $\chi^2$ value, it does so with additional small surface brightness features not present in the input map. This suggests that observations spanning multiple rotation cycles would be helpful.

\begin{table}
\caption{Summary of input and best fit values for the \textsc{PIMMS} code when running on the minimum required number of observations, equally spread across a single rotation period, of the simulated $\beta$ Cep and SPB stars described in Sect. \ref{obs_req}. The number of observations is 19 and 109 respectively. The $\beta$~Cep star has 3 pulsation modes, and spots on the magnetic poles, while the SPB has a single pulsation mode and off-pole spots. Both models have dipolar magnetic fields.}
\label{table:4 - equal obs results} 
\centering
\begin{tabular}{c c c c c}
\hline\hline       
 & \multicolumn{2}{c}{$\beta$~Cep model} & \multicolumn{2}{c}{SPB model}\\
 Parameter &Input&\textsc{PIMMS}&Input&\textsc{PIMMS}\\
\hline                                   
    $v_{p,1}$ & 20.0000 & 20.0194 & 15.0000 & 14.1673\\
    $\varphi_1$ & 0.0000 & 0.0005 & 0.2000 & 0.2004\\
    $K_1$ & 0.0000 & 0.0005 & 0.0100 & 0.0129\\
    $v_{p,2}$ & 0.5000 & 0.5074\\
    $\varphi_2$ & 0.3000 & 0.3002\\
    $K_2$ & 0.0100 & 0.0110\\
    $v_{p,3}$ & 0.5000 & 0.5205\\
    $\varphi_3$ & 0.7000 & 0.7091\\
    $K_3$ & 0.1000 & 0.0926\\
    $B_\text{d}$ & 583 & 581 & 583 & 589\\
\hline                                             %inserts single line
\end{tabular}
\end{table}

Now we consider observations scheduled over 6 months (to reflect a telescope semester of observations) with the gaps in phase over both rotation and pulsation cycles minimised. For our model $\beta$ Cep pulsator, this only slightly improves the minimum number of observations to 18. By improving the phase coverage in pulsation, a marginally better fit is obtained in the pulsation step with less data required. For our model SPB, the required number of observations is 52. Spreading the observations across multiple rotation cycles drives this reduction (see Sect. \ref{puls_rot_ratio}). 

\begin{table}
\caption{Summary of input and best fit values for the \textsc{PIMMS} code when running on the minimum required number of observations, where the gaps in rotation and pulsation phase are minimised, of the simulated $\beta$ Cep and SPB stars described in Sect. \ref{obs_req}. The number of observations is 18 and 52 respectively. The $\beta$~Cep star has 3 pulsation modes, and spots on the magnetic poles, while the SPB has a single pulsation mode and off-pole spots. Both models have dipolar magnetic fields.}

\label{table:5 - optimal obs results} 
\centering
\begin{tabular}{c c c c c}
\hline\hline       
 & \multicolumn{2}{c}{$\beta$~Cep model} & \multicolumn{2}{c}{SPB model}\\
 Parameter &Input&\textsc{PIMMS}&Input&\textsc{PIMMS}\\
\hline                                   
    $v_{p,1}$ & 20.0000 & 20.0194 & 15.0000 & 14.8760\\
    $\varphi_1$ & 0.0000 & 0.0005 & 0.2000 & 0.2004\\
    $K_1$ & 0.0000 & 0.0005 & 0.0100 & 0.0116\\
    $v_{p,2}$ & 0.5000 & 0.5074\\
    $\varphi_2$ & 0.3000 & 0.3002\\
    $K_2$ & 0.0100 & 0.0110\\
    $v_{p,3}$ & 0.5000 & 0.5205\\
    $\varphi_3$ & 0.7000 & 0.7090\\
    $K_3$ & 0.1000 & 0.0950\\
    $B_\text{d}$ & 583 & 581 & 583 & 586\\
\hline                                             %inserts single line
\end{tabular}
\end{table}

Finally we consider observations scheduled at random over 6 months, and repeat this test 100 times. This situation occurs when the stellar rotation period is not known before observations or when phase-constrained observations cannot be scheduled at the telescope. We find that the average minimum number of observations required to fit the model $\beta$ Cep with $\chi^2<1.5$ is 24, while the model SPB is 68. As a general rule, gaps in rotation phase over 0.15, and in pulsation phase over 0.25, will not provide a well constrained magnetic map and pulsation parameters. We stress that these numbers are for simple examples with few low-order pulsation modes, and a dipolar field. More complex stars will require more observations, as discussed below. 

\section{Considerations for application to real stars}\label{discussion}

\subsection{Limits of Zeeman-Doppler Imaging as a technique}

First, it is important to briefly discuss the limits of Zeeman-Doppler Imaging itself, such that the distinction can be made of whether deficiencies occur in just \textsc{PIMMS} or the technique as a whole. ZDI is a tomographic technique and hence is dependent on rotational modulation detectable in the observations. For stars with low $v_\text{eq}\sin i$ values, high resolution spectropolarimetry is necessary to be able to detect variations in the line profile. In the case of stars that are pole-on, there is no modulation in the line profiles associated with rotation, and therefore it is impossible to perform ZDI, regardless of the code used. Likewise, for equator-on stars, there is a degeneracy between the North/South hemispheres which makes ZDI mapping inevitably inaccurate. In fact, with Stokes V inversion only there is an overall uniqueness problem with ZDI, well exemplified by the study of $\tau$~Sco by \cite{uniqueness_problem}. Multiple different magnetic field strengths and geometries produced indistinguishable Stokes V profiles for this star. It is also well established in ZDI that there is significant crosstalk between the radial to meridional magnetic fields at low latitudes (\citealt{crosstalk}; \citealt{olegs_numerical_tests}). While it is possible to largely solve the uniqueness problem by introducing Stokes Q and U observations, this is impractical for many stars due to the significantly higher S/N required for a magnetic detection in linear polarisation data. Regardless of the Stokes parameters used, fine scale structure will always be lost due to its sensitivity to noise and intrinsic line broadening, which provides a limit on the spacial scales that can be reconstructed.

It is necessary to use Stokes I data to reconstruct a map of surface weights to the measured magnetic field \citep[e.g.][]{temperature_spots}. These can be in the form of brightness, temperature or abundance maps. Spots recovered in ZDI are often smeared due to the sampling of the data, like we found in Sect. \ref{spot_sec}. This has little effect on the recovered magnetic field if not subject to over-fitting.

\subsection{Model resolution requirements}

\begin{figure}
  \resizebox{\hsize}{!}{\includegraphics{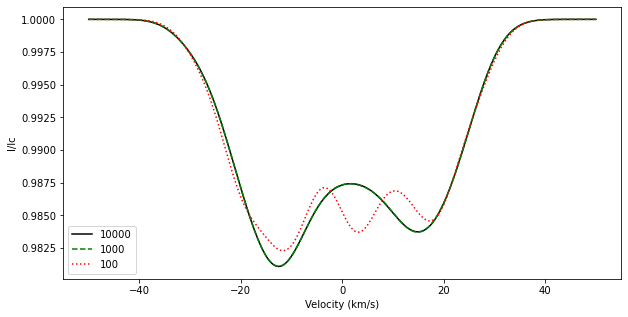}}
  \caption{A comparison between the line profiles of a $\ell=5$, $m=-5$ mode generated for three different resolutions. The solid black line is the higher resolution model using 10000 equally sized cells on the stellar surface, while the green dashed line is the 1000 cell moderate resolution model, and the red dotted line is the lower resolution model using only 100. The agreement between the 10000 and 1000 cell models show that the shape of the 100 cell model is incorrect due to an insuffucient resolution.}
  \label{resolution_comparison}
\end{figure}

Care must be taken when selecting the number of surface elements used for the model star. While running with a unnecessarily high resolution does not produce negative outputted results, it does strongly affect the runtime. On the other hand, an insufficient resolution can significantly harm the results. Higher degree pulsation modes cause sharper changes in velocity, which cannot be well represented in a low resolution map. High velocities also cause an increase in the difference in  Doppler shift between surface elements, so a greater spacial resolution is needed for stars with large pulsation velocities, or $v_\text{eq}\sin i$ values. Additionally, low resolution increases the level of smoothing in the brightness maps, therefore affecting strengths in the magnetic map. Correct selection of the resolution can be verified by looking at the outputted model line profiles, e.g. Fig.~\ref{resolution_comparison}. 

\subsection{Signal-to-noise requirements}

The tests shown in Sect. \ref{test} use data with a signal-to-noise ratio (S/N) in the model line profile of 1900, but we have successfully mapped model stars with S/N down to 250. This would be sufficient to detect a strong magnetic field in a star with a low $v_\text{eq}\sin i$, e.g. the magnetic fields of B stars which have a median $B_\text{d}$ value of 3~kG \citep{Shultz}. Added noise has more significant effects on fitting the brightness map than the magnetic field mapping individually, but when fitting both together the noise can create errors in field structure and strengths. Ultimately, the limiting factor for S/N is more likely to be in detecting the magnetic field in the first place, and pulsation mode identification if done with spectropolarimetric data.

\subsection{Pulsation phase coverage}

As briefly mentioned in Sect. \ref{obs_req}, the phase coverage requirement for pulsation is less than that for rotation. Like all forms of ZDI, the rotation phase must be covered as we cannot map the surface of a star without seeing the full surface, nor recover locations of features without good indications of when they rotate in and out of view. In contrast, the pulsation phase does not need to be as well covered in the case where the modes have already been identified, as all we are attempting to fit here is the phase, tangential to radial displacement ratio, and velocity amplitude for each mode. This is simpler than full mode identification.

\subsection{Rotation-pulsation period ratio}\label{puls_rot_ratio}

In some stars the rotation and pulsation frequencies can be very similar, e.g. SPBs commonly have both rotation and pulsation periods of the order of 1~c/d. If data is taken over a short baseline, at every point in the rotation phase observed, the same point in pulsation phase will be sampled. This makes it difficult to disentangle the pulsations from permanent features. When scheduling observations for such cases, the base line of observations must be considered. It is best practice to ensure the observations are taken over an entire semester, rather than during a short stint. 

\subsection{Dependence on the pulsation modes}

The higher the number of modes, the more data that is required to fit the pulsation parameters. This is independent of whether or not the modes are identified in advance of the spectropolarimetric observations being taken. Likewise, this is the case for more complex mode geometries, especially when there are multiple pulsation modes present. With fewer observations, and poorer phase coverage, the code is more likely to confuse features from different modes causing significant errors in the brightness and magnetic field mapping. 
Errors in mode identification can also cause significant issues, though the errors in this case tend to lead to obviously incorrect results, e.g. extreme bright/dark spots in the brightness map similar to those seen while performing ZDI on pulsating stars with \textsc{ZDIpy}. In some cases, this could mean that the brightness mapping step could help improve mode identification.

\subsection{Pulsation amplitudes}

If line profile variations due to pulsation are not detectable for a star, it is unlikely that \textsc{PIMMS} is required to map them. Past the detection threshold, it is ultimately a question of the required precision for the desired use. In the case of performing magneto-asteroseismology, we are striving for maximal precision in magnetic field strength and geometry, which requires use of \textsc{PIMMS}. In the case of simply estimating magnetic field strengths and geometries of stars with small pulsation amplitudes, the increased runtime and required pulsation mode identification can be skipped for a reduction in precision.

\section{Discussion and Conclusions}

In this paper we have shown for the first time that Zeeman-Doppler Imaging can be performed accurately on pulsating magnetic stars by including an additional time-dependent velocity map to the fitting procedure. \textsc{PIMMS} will allow us to obtain a more precise characterisation of the magnetic field at the surface of pulsating stars. This will be essential for constraining magneto-asteroseismic models in the future. The upcoming PLATO mission, scheduled to launch at the end of 2026, will provide long baseline, high cadence photometric data \citep{PLATO}. \textsc{PIMMS} will be able to contribute to the ultra-precise stellar characterisations needed both for PLATO exoplanet and stellar evolution research.

The results we obtain with our code \textsc{PIMMS} are subject to the same limits as standard ZDI faces. \textsc{PIMMS} also has a larger data requirement due to the pulsation phases that must be additionally covered and the velocity map that must be fitted. The magnitude of the increase in data is dependent on the number of pulsation modes and their geometries, as well as observing strategies for sampling the rotation and pulsation phases. This makes it hard to predict the number of spectropolarimetric observations needed without first characterising the pulsations.

The model used in this version of \textsc{PIMMS} only considers the effects of pulsation on velocity, while it is known to also affect temperature/brightness. For slowly rotating SPB and $\beta$~Cep stars, \cite{temp_var} has shown the effect of local temperature variations on the surface-integrated line profiles are negligible. In the case of rapidly rotating stars, the spherically symmetric description of the star and its pulsations breaks down, so the code cannot accurately model such a star's line profile variations regardless. Further work to include the effects of rapid rotation into the stellar models are not currently planned due to the difficulty in detecting their magnetic fields with spectropolarimetry, and the complexity in identifying their pulsation modes. 

It is also currently assumed in \textsc{PIMMS} that the line forming region is geometrically thin, such that the local Doppler shift is not dependent on line depth. Zeeman-Doppler Imaging is commonly performed on average line profiles calculated using the least-squares deconvolution technique (LSD; \citealt{lsd}), which uses the same assumption, so is consistent. Stellar atmospheres are however stratified by magnetic fields, as they strongly influence diffusion. This could allow us to probe different layers of the stellar atmosphere \citep{line_forming_depth}. Incorporating such depth-dependence into the model would require a large number of free parameters, and reduce the SNR of the data as LSD would no longer be possible, hence should be approached with caution and thoroughly tested in the future.

Currently, \textsc{PIMMS} requires that the degree $\ell$, azimuthal order $m$, and frequency $f$ of each pulsation mode is provided a priori, i.e. mode identification must be performed with another method. As it is possible to perform mode identification from the spectropolarimetric data itself, and it is known that magnetic fields affect the allowed pulsation modes, it would therefore be better if the mode identification was performed simultaneously with the magnetic field mapping. Using Doppler shifts to map pulsations has been done previously by \cite{doppler_pulsation_code}, and allows for a more complex description without a parameterisation using a small number of specific spherical harmonics.

The next step is to begin applying \textsc{PIMMS} to real stars. 
A few magnetic pulsating stars with a sufficient spectropolarimetric data set, and well characterised pulsation modes, have been identified and will be the subject of future papers. In addition, we are currently performing an observing campaign to gain sufficient data for additional magnetic pulsating targets. 

\begin{acknowledgements}

We thank the anonymous reviewer for their constructive comments, which in particular helped improve the clarity of Sect. \ref{stokes_sec}. We are also grateful to Oleg Kochukhov for valuable discussions throughout the development of this software. CPF received funding from the European Union's Horizon Europe research and innovation programme under grant agreement No. 101079231 (EXOHOST), and from the United Kingdom Research and Innovation (UKRI) Horizon Europe Guarantee Scheme (grant number 10051045). 

\end{acknowledgements}

\bibliographystyle{aa} % style aa.bst
\bibliography{main}

\end{document}